\documentclass[namedreferences,hyperref,optionalrh]{spr-sola}
\usepackage{graphicx}        
\usepackage{color}           
\usepackage{soul}
\usepackage{todonotes}
\usepackage{hhline}
\chardef\us=`\_

\def\er{ER}
\def\ers{ERs}
\newcommand{\Fig}[1]{Figure~\ref{#1}}

\begin{document}

\begin{frontmatter}
\title{Large ephemeral regions and their tilt angles}

\author[addressref={aff1},email={rambahadurgupta.phy23@itbhu.ac.in}]{\inits{R.}\fnm{Rambahadur}~\snm{Gupta}\orcid{0009-0003-7500-7258}}
\author[addressref=aff1,email={anubsreedevi.rs.phy20@itbhu.ac.in}]{\inits{A.}\fnm{Anu}~\snm{Sreedevi}\orcid{0000-0001-7036-2902}}
\author[addressref=aff2,email={bkjha.sun@gmail.com}]{\inits{B.K.}\fnm{Bibhuti~Kumar}~\snm{Jha}\orcid{0000-0003-3191-4625}}

\author[addressref={aff1}, corref,email={karak.phy@iitbhu.ac.in}]{\inits{B.B.}\fnm{Bidya~Binay}~\snm{Karak}\orcid{0000-0002-8883-3562}}
\address[id=aff1]{Department of Physics, Indian Institute of Technology(Banaras Hindu University), Varanasi 221005, India}
\address[id=aff2]{Southwest Research Institute, Boulder, CO 80302, USA}

\runningauthor{Gupta et al.}
\runningtitle{A study of ephemeral regions}

\begin{abstract}

The ephemeral regions (ERs), which are short-lived bipolar magnetic regions that emerge across the solar cycle but do not appear as sunspots, play a crucial role in the Sun's magnetic flux budget. However, their properties, particularly the tilt distribution, are poorly constrained by observations.   In this study, we isolate ERs from the Automatic Tracking Algorithm for Bipolar Magnetic Regions (AutoTAB) catalog during Solar Cycles 24 and 25 by applying flux and footpoint-separation thresholds.   Although AutoTAB was designed to track high flux regions, it also records ephemeral regions (ERs) with fluxes of $10^{19}$–$10^{20}$\,Mx, placing them at the upper end of the ER spectrum. The isolated ERs have an average lifetime of 1.2\,days. Footpoint separation begins at supergranular scales ($\sim 20$\,Mm), grows during the first half of the lifetime, and then saturates.
 
ERs occur most frequently near solar minima, consistent with earlier studies and likely reflecting AutoTAB's greater sensitivity to weaker regions when strong BMRs are scarce. Tilt properties reveal a more complex picture. For lifetimes shorter than two days, ERs show a broad, noisy distribution with no systematic latitude dependence. Including longer-lived ERs produces a weak, though statistically insignificant, increasing trend in latitude, suggesting that short-lived ERs are shaped by turbulent convection, while stronger, longer-lived ERs may retain Coriolis-imparted tilts. Overall, these results support the view that ERs occupy the low-flux end of the BMR spectrum and contribute meaningfully to the solar dynamo.
\end{abstract}
\end{frontmatter}

\section{Introduction}
     \label{Intro} 

The solar surface harbors magnetic features that have diverse shapes and sizes, spanning a wide range of fluxes. The most prominent and noticeable of these are the active regions or sunspot pairs. These regions appear as dark features on the solar photosphere in white light observations due to their lower temperature caused by the presence of strong magnetic fields \citep[for comprehensive review on this topic, see][and references therein]{Solanki2003}. On a closer look, these sunspot pairs observed correspond to regions of strong and opposite polarities located in close proximity in magnetic field observations. These structures are commonly referred to as Bipolar Magnetic Regions \citep[BMRs:][]{H1991, H1996}. Frequency of occurrence of such BMRs varies cyclically over the $\approx$11\,years, emerging at mid latitudes at the beginning of the cycle and slowly migrating equatorward with the progress of the cycle \citep{S1844,M03}. The orientation of emerging BMRs typically follows a systematic pattern, with the leading polarity situated closer to the equator than the following polarity--Hale’s polarity law \citep{HN1925}. This leads to an apparent angle between the equator and the line connecting the two polarities, which statistically increases with latitude--Joy’s law \citep[]{HE1919,WS89}.

Although sunspots represent the higher end of the magnetic structures that we observe, many bipolar structures possess magnetic fields too weak to generate sufficient contrast to be detected in the continuum observations. A subset of such BMRs usually tends to have even shorter lifespans and smaller fluxes than an average active region. This group is popularly known as Ephemeral Regions (\ers) \citep{HM1973, LG2015}. As they persistently appear at all times and in large numbers throughout the solar disk, \ers\ and their role in solar dynamo have been studied from the advent of magnetic observations \citep{Harvey1975, HM1973, HD2008, HK2024}. Consequently, it is seen that the definitions of \ers\ vary throughout the literature, for example,  \cite{Hagenaar2001} defined \ers\ as regions with flux ranging from $10^{18}$ Mx to $3 \times 10^{20}$ Mx, while \cite{HM1973} used a flux range of $10^{19}$ Mx to $4.7 \times 10^{20}$ Mx.

Unlike the behavior expected for BMRs, where their emergence closely follows the solar cycle, previous studies found no clear solar cycle type variation in \ers   \cite{HM1973, TP2011}. Studies conclude that these regions tend to emerge predominantly near the supergranular boundaries \citep{Hagenaar2001, HK2024} and at a higher degree of random orientation compared to BMRs; only about 60\% of \ers\ conform to Hale’s polarity law \citep{Hagenaar2001, Hagenaar2003}. 
\citet{HM1973} reported that the emergence rate of \ers\ is higher than the BMRs with an estimated 100 new \ers\ emerge per day across the solar disk \citep{HM1973}. Furthermore, the flux emergence rate due to \ers\ is estimated to be as high as $2.8 \times 10^{23}$ Mx per day \citep{chae2001}.
This suggests that \ers\ are responsible for a substantial magnetic flux input into the photosphere and may contribute to the heating of the upper atmosphere \citep{Schrijver1998, smartin_review}.

The physical origin of \ers\ remains an open question and is the subject of ongoing debate. Several studies proposed that \ers\ are generated by a small-scale dynamo process operating near the solar surface 
\citep{N1992, C1999, Hagenaar2003, Yang2014}, while other studies suggest that they could represent fragmented remnants of decaying active regions \citep{Hagenaar2008} and thus are essentially part of the large-scale dynamo \citep{Cha20, HK2024}. 

\ers\ may not just simply be the byproducts of large-scale dynamo, they may contribute to it if they have a statistically non-zero tilt, despite having smaller temporal and spatial scales than BMRs. Due to their finite tilt, bipolar regions decay to generate a large-scale  poloidal field that supports the solar dynamo \citep{L1964, CS15, WYM18, K2023}.

This raises a natural question: How much is the contribution of \ers\ to the polarity reversal, and in what way do they contribute to the stochasticity in solar cycles which are believed to be primarily caused by the scatter of the BMR properties \citep{KM17,Kumar24}? Although there is strong evidence that \ers\ emerge with random orientations, their high emergence rate (and thus the flux) compared to the active regions, even a weak non-zero tilt may contribute to a significant component in the poloidal field budget and thus in the global dynamo \citep{TG2010,TP2011}.

In the present work, we focus on the tilt angle of ephemeral regions, examining whether they exhibit a statistically significant tilt and whether they follow the trend prescribed by Joy’s law. For this purpose, we use the catalog of ephemeral regions detected and tracked with our recently developed Automatic Tracking Algorithm for BMRs (AutoTAB; \citealt{JP2021, SJ2023}).
 
A novelty of our work is that our \ers\ are tracked during their evolution on the near side of the solar disk, and thus they are not counted repeatedly, which otherwise results in higher weights for larger regions.

\section{Data and Methodology}
\subsection{Data}

The dataset used in this study is sourced from
AutoTAB which lists tracked information of $\sim 12,000$ BMRs during 1996-2023 generated using AutoTAB. The algorithm automatically detects and tracks BMRs, through their nearside evolution, using line-of-sight (LOS) magnetograms from Magnetic Doppler Imaging \citep[MDI: 1996–2011 at 96~minutes cadence][]{SB1995} and Helioseismic and Magnetic Imager \citep[HMI: 2010–present at 96~minute cadence][]{SchouS2012}) on board the Solar and Heliospheric Observatory (SOHO) and Solar Dynamic Observatory (SDO). The working of AutoTAB is summarized below.

AutoTAB operates in two stages: detection and tracking. The detection part follows the steps prescribed in \citet{SK2012} which was also adopted by \citet{JK2020} with slight modifications. The algorithm starts by correcting the line-of-sight (LOS) magnetograms for projection effects to derive the radial component of the magnetic field. It then identifies regions of strong magnetic flux in the magnetograms and applies a flux balance criterion to isolate bipolar regions along with a cutoff of absolute maximum recorded value in the BMR to be grater than 200~G. Once BMR is detected, they are saved as binary maps, which serves as input to the tracking part of AutoTAB. For optimized tracking, identified regions undergo pre-processing steps to extract approximate boundaries of the individual BMRs. This initiates the tracking part of the AutoTAB. A BMR is tracked by differentially rotating its isolated mask forward and checking for spatial overlap with masks from later times (see Sections 2.2 and 2.3 in \citet{SJ2023} for AutoTAB details). This process enables AutoTAB to follow the evolution of individual BMRs across multiple observations. Here we note that we loose the track of a BMR once it crosses to the farside.

Thus, the AutoTAB catalog comprises comprehensive tracked information of all BMRs across wide flux ranges ($10^{19}-10^{24}$~Mx), spanning from their first detection to their final stage on the Sun's near-side. Previously, \citet{SJ2024} focused exclusively on long-lived BMRs, removing short-lived \ers\ and considering only those with lifespans longer than 8 hours. In contrast, this study focuses on the tilt properties of short-lived BMRs (\ers), analyzing each at its mature phase. Most of the previous work which studies BMRs or \ers\ is done by treating all the tracked BMRs as new independent emergence events, but in our work we consider each \ers\ only once, ensuring a more physically meaningful representation of the tilt characteristic of BMRs. Each \er\ parameter is calculated as the mean over all instances where the total unsigned magnetic flux of the \er\ exceeds 80\% of its maximum, between $\pm55^{\circ}$ longitude. This method of assigning single representative value for a BMR has been previously used in \citet{SJ2024}. This criterion ensures that measurements are taken when BMRs are in the mature phase--- they acquire their maximum flux, and the error due to projection effects is less. 

\begin{figure}
\centering
\includegraphics[width=0.5\textwidth]{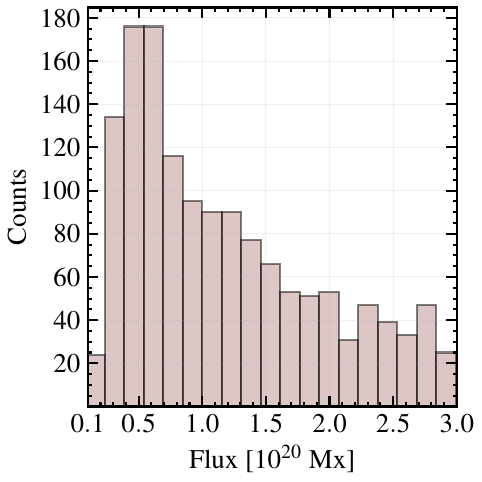}
\caption{Distribution of the (total unsigned) flux of \ers\ used in our study.}
\label{fig:FluxDistrib}
\end{figure}

To filter out weaker and less significant BMRs from the AutoTAB catalog, we applied an upper flux threshold $3 \times 10^{20}$ Mx, following the work of \citet{Hagenaar2003}, and an initial footpoint separation threshold of 40~Mm at their first detection, chosen in regard with the previous study of \citet{SJ2024} (their Figure~4 (b)) which shows the footpoint separation evolution for this particular flux range. We find that these thresholds efficiently filter out low flux BMRs in \er\ ranges from the AutoTAB catalog. Filtered out \er\ catalogue comprises of 1868 \ers. During solar cycle~23, we identify significantly lower number of \ers\ in AutoTAB catalog. This can be accounted for poor quality of MDI data, during that period compared to HMI observations in the later time. Hence, in this study, we restrict our analysis to BMRs from solar cycles 24 and 25. The flux distribution of the filtered data, which we refer to as \ers, is shown in \Fig{fig:FluxDistrib}. As expected, they show a log-normal type distribution and the number of detected \ers\ increases with the decrease of flux. 
As we lower the upper flux threshold, fewer \ers\ gets filtered out. Eventually, the count drops sharply to zero because AutoTAB detects strong flux regions using a threshold based on the average magnetic flux from the magnetogram. This imposes a natural lower limit on the number of \ers\ detected. As we are interested in studying the tilt angle of moderate and high-flux \ers, removal of small-flux \ers\ (with flux $\le 10^{19}$~Mx), which have random orientations \citep{Hagenaar2003}, becomes an advantage for us. We additionally note that, in this study, the tilt is computed following the standard formulation proposed by \citet{WS1991} using the flux-weighted centroid of the polarities.

\section{Results}

\begin{figure}
\centering
\includegraphics[width=0.7\textwidth]{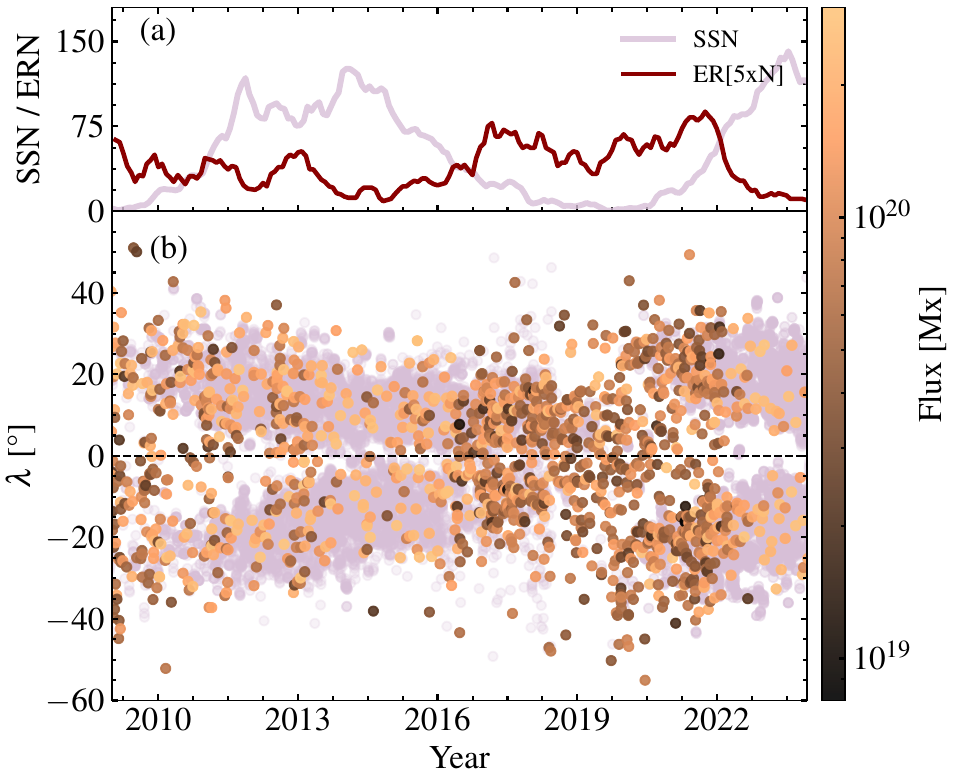}
\caption{
(a) Variation of monthly \er\ number identified in AutoTAB in red dark with the traditional monthly sunspot number (ISSN V2.0) in light color for reference. (b) Latitude–time (butterfly diagram) distribution of tracked \ers.
Total unsigned magnetic flux in each identified region is represented by the color gradient (see colorbar). 
Points with light color in the background show the positions of sunspot groups obtained from \citet{Mandal2020}.
}
\label{fig:butterfly_Comb}
\end{figure}

\subsection{Time evolution of \ers}
Having discussed their flux distribution, we now examine how the occurrence of \ers\ varies over the solar cycle. \Fig{fig:butterfly_Comb}(a) presents the appropriately scaled number of \er\ detected over two solar cycles, together with the monthly averaged sunspot number (shown in light color) for comparison. The distribution of the detected \ers\ in latitude over their emergence time is shown in \Fig{fig:butterfly_Comb}(b). It is important to remember that during solar maxima, our adaptive thresholding (based on the average magnetic flux of each magnetogram) tends to exclude more 
\ers\ from detection. Despite this limitation, the \ers\ conform to the expected latitudinal evolution, though not that strictly: at the beginning of each cycle, \ers\ emerge at higher latitudes and gradually migrate towards the equator as the cycle progresses. They generally confirms to the expected latitude bands of the butterfly wing-like pattern but exhibit a broader latitudinal spread compared to the original BMRs. Notably, during the two solar minima (around 2008 and 2020), we detect more \ers\ than during solar maxima, with a particularly high density observed around the 2020 (solar cycle 25 minimum). Interestingly, most of the high-flux \ers\ observed in our study are also concentrated during these minima. While this trend largely reflects the sensitivity of our detection algorithm, previous studies have also reported an increased presence of \ers\ during solar minimum \citep{TG2010}, possibly linked with the enhancement of the generation of low-flux magnetic structures through the small-scale dynamo due to the reduced large-scale magnetic field during minima, as found in dynamo simulations of \citet{KB2016}.

\begin{figure*}
\centering
\includegraphics[width=\textwidth]{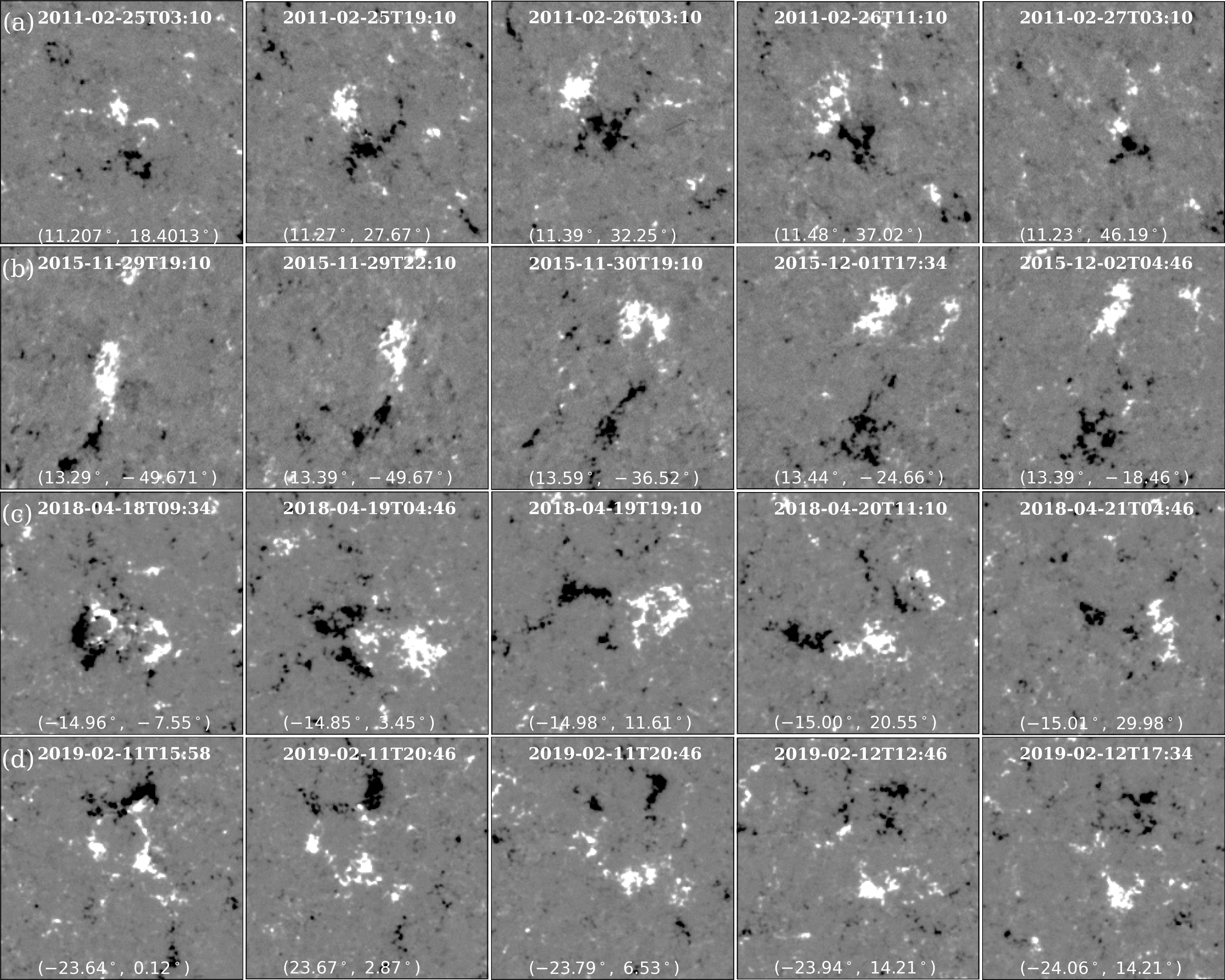}
\caption{Panels (a) and (d) show the snapshots of the tracked \ers. The LOS magnetic field is saturated at $\pm$ 100~G.}
\label{fig:typicalEARs}
\end{figure*}

The average lifetime of the tracked \ers\ in our dataset is approximately 1.2~days. Owing to their relatively low magnetic flux and short-lived nature, tracking such regions poses significant observational challenges. According to the classical $\Omega$-loop model for the BMR formation, the initial emergence phase is crucial. However, for low-flux regions like \ers, this phase is often missed in LOS magnetograms because the magnetic field is initially horizontal and weak. Only as the loops rise and the field becomes more vertical (radial), it becomes detectable in LOS observations. As a result, many of the \ers\ we capture are already in their mature or decaying stages when they first become detectable 
\citep[however see,][for a through study capturing the emergence phase of ERs visually]{Yang2014}. With this limitation in mind, we examine the time evolution of a few typical \ers, as shown in \Fig{fig:typicalEARs}. These examples reveal that their magnetic structures often deviate substantially from the idealized bipole configuration \citep{Y2020}, despite enforcing a minimum 40\% flux balance between the positive and negative polarities during detection. Furthermore, they exhibit highly dynamic behavior, with rapid morphological changes observed throughout their tracked lifetimes \citep{Yang2012}. 

\begin{figure}
\centering
\includegraphics[width=1\textwidth]{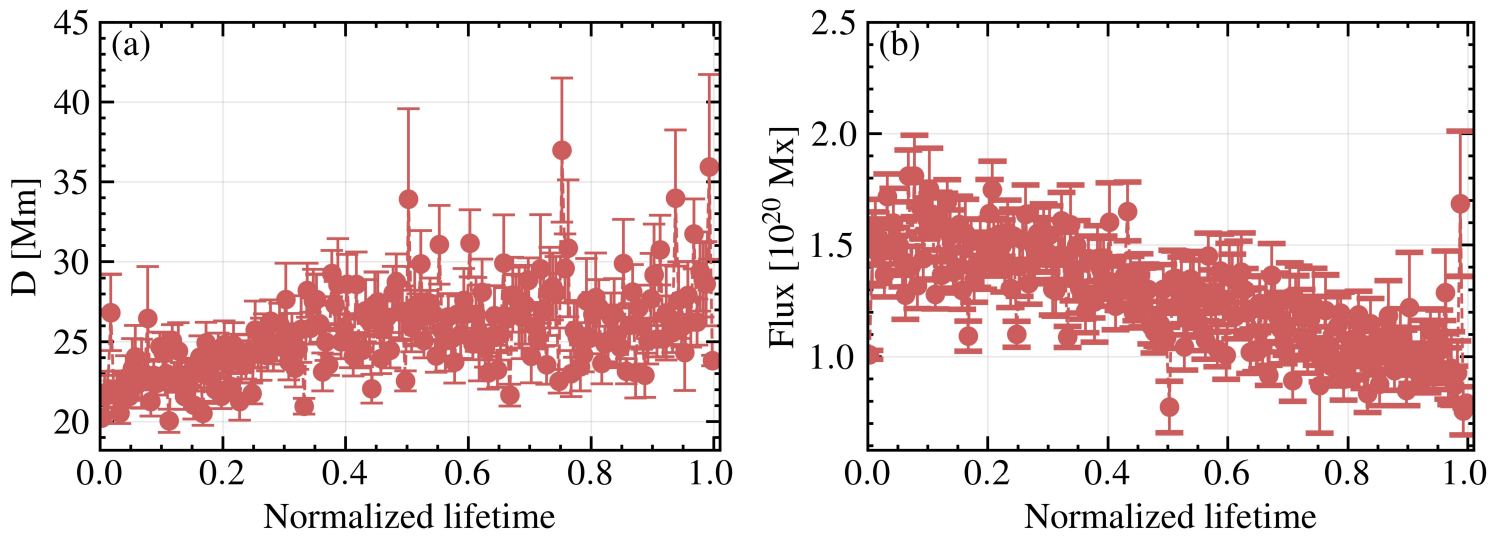}
\caption{Collective footpoint separation ($D$) and unsigned flux ($\Phi_m$)  evolution of ERs during their normalized lifetime.}
\label{fig:fp_fx}
\end{figure}

These expected behaviors from \er\ evolution are clearly reflected in how their footpoint separation and unsugned flux change over time. In \Fig{fig:fp_fx}, we plot the footpoint separation and flux of all \ers\ by normalizing the detection time of each \er\ with its maximum detected time with the data averaged over 0.005 binsize and the standard error of the average is shown as the error bar. The separation between the two poles of each \er\ starts at around 20~Mm, at supergranulation length scales and increases steadily until about 50\% of their tracked lifetime. Beyond this point, the separation appears to saturate.

This trend suggests that, over the course of their observed lifetimes, ERs evolve on spatial scales comparable to those of supergranulation, although the initial emergence phase is not resolved in our data.

Unlike the larger BMRs studied in \citet{SJ2024}, the decaying phase of small regions such as \ers \ is captured more efficiently by AutoTAB and is evident as the 
unsigned flux smoothly decreases in time (\Fig{fig:fp_fx}(b)). 

The large scatter observed in footpoint separation in the later stages is likely due to the increasingly fragmented decay of these small regions and fewer \ers\ in these bins.

\subsection{Tilt angle and Joy's law}

\begin{figure}
\centering
\includegraphics[width=1\textwidth]{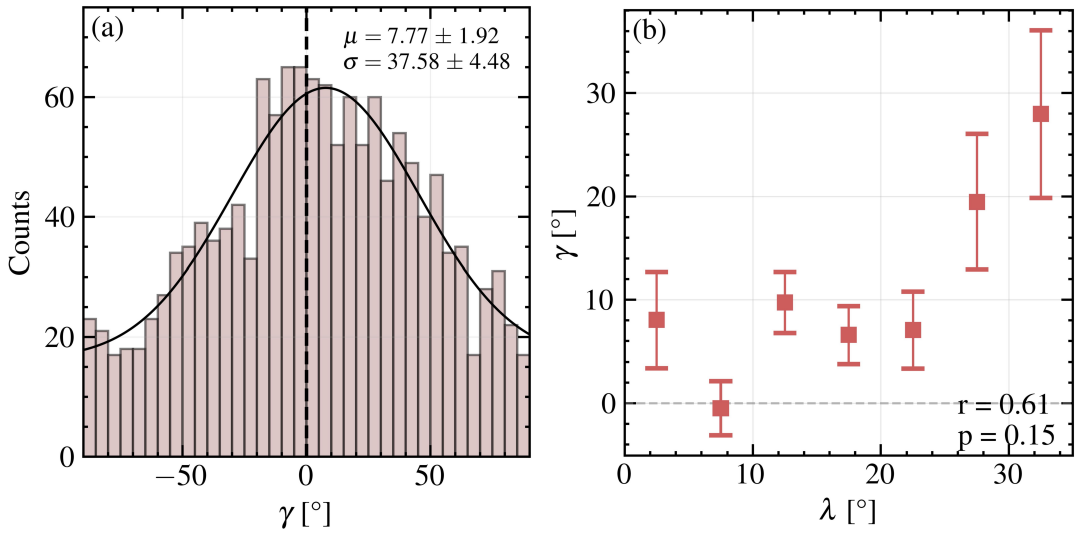}
\caption{(a) Tilt distribution: number of BMRs in $5^\circ$ tilt bins are shown as bars. The solid line represents the Gaussian ﬁtted curve (with an offset) with ﬁtting parameters mentioned in the panel. The vertical solid blue line represents the 0° tilt. (b) Fitted Gaussian mean tilt in each $5^\circ$ latitude bin as a function of the latitude (Joy's law plot). The grey dashed line represents the $0^\circ$ tilt line. }
\label{fig:tilt_Distribut}
\end{figure}

The distribution of the tilts of \ers\ is shown in \Fig{fig:tilt_Distribut}(a). 
As apparent in the plot, the tilt distribution is broad, with a Gaussian fit centered at $7.77^\circ$$\pm1.92^{\circ}$ and a spread of $37.58^\circ$$\pm4.48^{\circ}$. 
While the tilt distribution itself appears scattered, it still suggests that \ers\ tend to emerge with a preference for positive mean tilt with about 54\% \ers\ are having positive tilts. It is these \ers\ which give correct polar field in the Babcock-Leighton process. In terms of the Hale's polarity law, about $72\%$ of total \ers\ obey this law.

\begin{figure}
\centering
\includegraphics[width=0.5\textwidth]{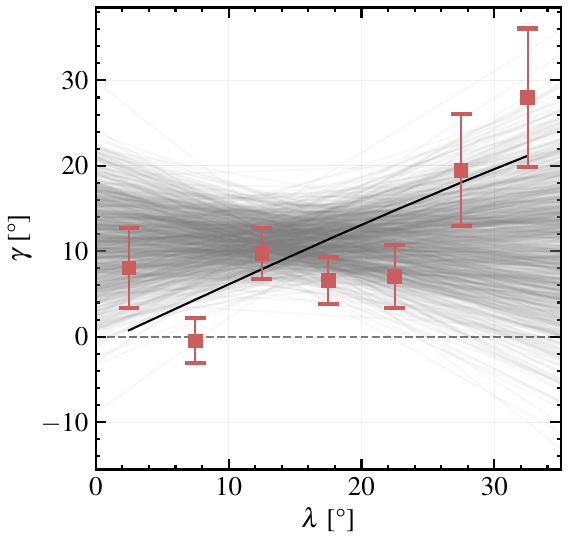}
\caption{
Red points with error: The observed tilt angles of \ers\ as a function of latitude (same as \Fig{fig:tilt_Distribut}(b)). These are now fitted with Joy's law: $\gamma = \gamma_0 \sin\lambda + b$, where $\gamma_0$ and $b$ are the fitted slope and the intercept, respectively, shown by the solid black line.
Gray lines are compared from the  random tilt realizations generated using Gaussian distributions centered at zero with standard deviations equal to the observed values (gray lines) in each latitude bin. All the plots are lifted up by the mean of observed data ($11.2^\circ$) to facilitate comparison.}
\label{fig:random_tilt}
\end{figure}

Next, we group the tilt measurements into $5^\circ$ latitude bins and compute the mean tilt and its uncertainty in each bin by fitting Gaussian functions to the distributions. These binned mean tilt values, along with their uncertainties, are shown in \Fig{fig:tilt_Distribut}(b). The latitudinal dependence is weak but clearly shows an increasing trend of tilt with latitude, having $r$-value of 0.61 ($p=0.15$) and $\gamma_0 = 41.37 \pm 14.86$
(where $\gamma_0$ is the slope of Joy's law appearing as: $\gamma = \gamma_0 \sin\lambda + b$). This trend is in agreement with the previously reported results of \ers\  \citep{HA1993} where we see a weak but not statistically significant latitudinal dependence on tilt. Nevertheless, the observed latitudinal variation is overwhelmed by random fluctuations. We verified this by generating random tilt values in each latitude bin using Gaussian distributions centered at zero and with standard deviations equal to those obtained from the observed tilts from \Fig{fig:tilt_Distribut}(b), with an additional shift of $11.2^\circ$ (mean of the observed binned data) to capture the nonzero mean of real data into the generated synthetic data.  As seen in \Fig{fig:random_tilt}, the test confirms that the weak Joy’s law signal observed in the data may not be genuine because of the large $\sigma$ in the original data which makes the observed curve  indistinguishable from random background variations.

\begin{figure}
\centering
\includegraphics[width=0.5\textwidth]{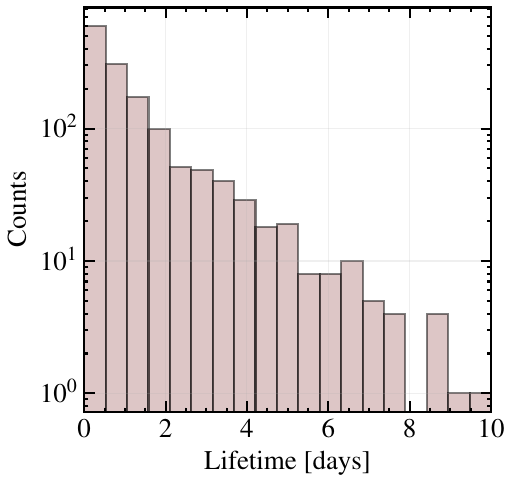}
\caption{Distribution of lifetimes of ERs detected and tracked using the AutoTAB. We note that the lifetime presented here is not the true lifetime of ER as AutoTAB often misses a part of the early emergence and the decay.}
\label{fig:lifetime}
\end{figure}

Up to this point, our selection of \ers\ was solely based on their unsigned flux and the footpoint separation. However, previous studies 
\citep[e.g.,][]{Harvey1975, Hagenaar2003}

have defined \ers\ based on their low unsigned flux and their shorter lifetimes. To understand how our \er\ population compares with this criterion, we show the lifetime distribution of AutoTAB-detected \ers\ in \Fig{fig:lifetime}. We find a mean lifetime of 29 hours and a median lifetime of 16 hours. Previously, \citet{HM1973} reported that \ers\ generally live less than 2 days, while \citet{Harvey1975} found a median lifetime of about 12 hours. These differences can largely be attributed to variations in data quality, cadence, and detection methodology. In  \Fig{fig:lifetime}), we see that a part of the \ers\ in the dataset persists well beyond the 2-day mark, highlighting a broader diversity in their temporal evolution. Hence, to facilitate a direct comparison with previous definitions, we additionally apply an upper threshold on the lifetime of \ers\ of 2 days to our \er\ sample, refining our selection to include only short-lived events. This shortlists 1164 \ers\ from our original dataset.

\begin{figure}
\centering
\includegraphics[width=1\textwidth]{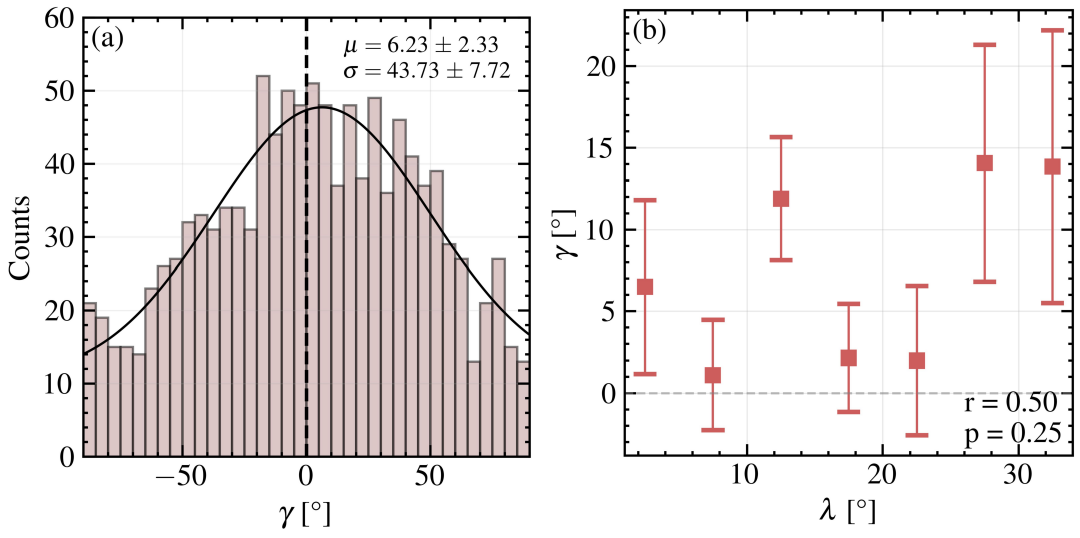}
\caption{The same as \Fig{fig:tilt_Distribut} but for \ers\ having lifetime $\le$ 2 days.}
\label{fig:Joy law_2}
\end{figure}

\begin{figure}
\centering
\includegraphics[width=1\textwidth]{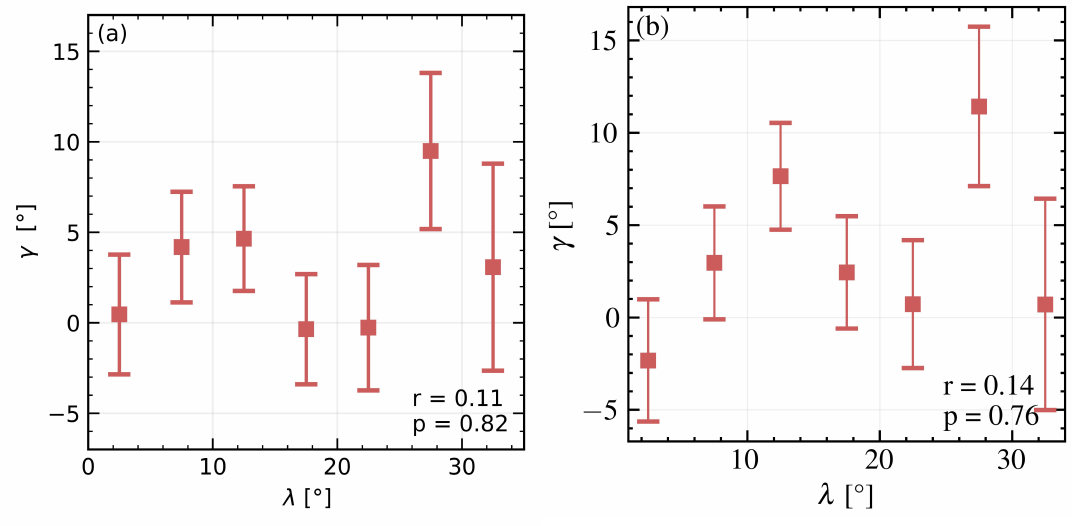}
\caption{Same as \Fig{fig:Joy law_2}(b) but with (a) arithmetic mean, and (b) median, instead of Gaussian fitted mean.}
\label{fig:Joy law_mean_med}
\end{figure}

Now, with this additional condition applied, we fit the results to the standard Joy’s law form and the resulting Joy’s law amplitude ($\gamma_0$) comes out to be $\gamma_0 = 16.15 \pm 12.70$ (see \Fig{fig:Joy law_2}).
A similar fit was performed using the median and mean tilt in each bin instead of the Gaussian mean, and subsequent $\gamma_0$ and $b$ are shown in Table~\ref{Tab:fit_parameters} and \Fig{fig:Joy law_mean_med}.
Even after a lifetime restriction imposed and comparing our results directly with the other studies, the tilt shows only a very weak latitude-dependent trend, as reflected from the large uncertainties of the sinusoidal fit and a low correlation coefficient of $0.5$ (with $p=0.25$), indicating no statistically significant dependence on latitude, similar to \Fig{fig:tilt_Distribut}(b).

\begin{table}
\begin{center}
\caption[]{Fit parameters of Joy’s law obtained from different fitting methods for every latitudinal bin. 
The columns list the slope ($\gamma_{0}$) and offset ($b$) of Joy’s law corresponding to each method. 
The asterisk (*) indicates that the ephemeral regions (ERs) are selected using only an upper flux threshold, 
whereas the other cases use a dataset additionally filtered by applying a lifetime threshold on top of the flux threshold.}
\label{Tab:fit_parameters}

\begin{tabular}{lcc}
\hline\noalign{\smallskip}
Fit in the Bin & $\gamma_{0}$ & $b$ \\
\hline\noalign{\smallskip}
Gaussian* & $41.37 \pm 14.86$ & $-1.06 \pm 5.04$ \\
Gaussian  & $16.15 \pm 1.50$  & $2.58 \pm 4.31$ \\
Mean      & $5.82 \pm 8.46$   & $1.31 \pm 2.87$ \\
Median    & $8.52 \pm 11.07$  & $0.84 \pm 3.76$ \\
\noalign{\smallskip}\hline
\end{tabular}

\end{center}
\end{table}

Upon comparing the Joy's law trends before and after applying the restriction on the lifetime of the \ers, we see that the trend becomes noticeably weaker,
with a lower correlation coefficient and a higher $p$-value, as shown in \Fig{fig:Joy law_2}. This suggests that some longer-lived regions currently classified as \ers\ may, in fact, be weak BMRs and mixing them in \er\ studies can corrupt the results. This highlights the need to reassess our classification criteria and develop a more refined definition of what constitutes an \er.

\section{Conclusion and Discussion}

This article investigates the properties of ERs extracted from the AutoTAB catalog by applying an upper threshold to the unsigned flux and lifetime. The average lifetime of the dataset is $\sim 1.2$ days. As is typical for BMRs, the flux distribution of \ers\ also follows a log-normal profile, consistent with earlier studies. The missing early emergence is evident in flux evolution plots, where the flux remains nearly constant in the initial phase of tracking before beginning to decay. Evolution of footpoint separation further supports this picture: starting at supergranular scales ($\sim 20$ Mm), the separation increases during the first half of the region’s lifetime and then saturates. 

Visibly, \ers\ are known to be present in large numbers at all times of solar cycle, all over the solar disk. Evaluating the latitude–time distribution of emerging \ers\ from our catalog reveals patterns that both agree and disagree with this: while we do trace the vague yet familiar butterfly pattern of sunspots with wider latitudinal distribution, the emergence of \ers\ shows a clear latitudinal dependence over the solar cycle. They start appearing at higher latitudes early in the cycle and migrate equatorward as the cycle progresses. This behavior, together with their bipolar structure and systematic emergence patterns, suggests that \ers\ may simply represent the low-flux end of the BMR spectrum rather than a distinct separate population. We also note that \ers\ are detected more frequently during solar minima. This may partly reflect AutoTAB’s bias, where weaker regions are more easily identified in the absence of strong BMRs, but it remains consistent with previous studies.

When \er\ emergences mimic BMR emergence patterns, their tilt properties appear to be more complex. The overall tilt distribution is broad and noisy, showing significant scatter and no statistical dependence on latitude.
However, when regions with longer lifetimes are included, 
the latitude dependence of the tilt is still not statistically significant, but the value of the mean tilt is increased. 
Large scatter in \ers\ tilt angles does not establish its root to Coriolis force-driven processes and may be attributed to turbulent convection. 
The stronger \ers\, with higher lifespan, is strong enough to 
resist convection and thus retain the imparted tilt in the CZ. 
These findings reinforce our previous conclusion: 
despite their small flux and large scatter around the mean tilt predicted by Joy’s law, the finite positive tilt of \ers\ (54\% as shown in \Fig{fig:tilt_Distribut}(a) with respect to $\sim 60\%$ for BMRs) indicates that it can make a measurable contribution to the reversal and building of the poloidal field in the Sun and recovery from extended episodes of inactive phases such as the Maunder minimum \citep{KM18, HK2024, smartin_review}. Furthermore, the conventional definition of \ers, based solely on flux or lifetime, may be insufficient and needs to be revisited to better capture their true physical characteristics in light of modern observations. Therefore, a more targeted analysis is warranted using specialized algorithms designed to detect and track \ers.
   
\section*{Acknowledgements}
B.B.K. acknowledges the financial support from the Anusandhan National Research Foundation (ANRF) through the MATRIC programme (file no. MTR/2023/000670) and the Alexander von Humboldt Foundation. The LOS magnetogram data used for the analyses in this article are obtained from HMI and downloaded via JSOC, courtesy of the NASA/SDO science teams. 
ISSN data are obtained from WDC-SILSO, Royal Observatory of Belgium, Brussels, DOI: \url{https://doi.org/10.24414/qnza-ac80} 
\citep{SILSO_Sunspot_Number}.
Sunspot area data were obtained from the calibrated sunspot dataset of \citet{Mandal2020}. 

\section*{Data Availability}
The AutoTAB catalog will be publicly available along with the codes at \url{https://github.com/sreedevi-anu/AutoTAB}. The analyses codes can be shared upon a reasonable request to the corresponding authors.

\bibliographystyle{spr-mp-sola}
\bibliography{ERs}


\label{lastpage}
\end{document}